\documentclass{article}
\usepackage{spconf}
\ninept

\usepackage{amsmath,graphicx}
\usepackage{cite}
\usepackage{graphicx}
\usepackage{psfrag}
\usepackage{epsfig}
\usepackage{url}
\usepackage{stfloats}
\usepackage{amsfonts,amssymb,amsmath,bm,paralist,theorem,cite,ifthen,color,amsmath}
\usepackage{array}
\usepackage{caption}
\usepackage{calc}
\usepackage{longtable}
\usepackage{subfigure}
\usepackage{mathrsfs}
\usepackage{epsfig}

\usepackage{algorithm}
\usepackage{algorithmic}
\theorembodyfont{\rmfamily}

\DeclareMathOperator*{\maximize}{maximize}
\DeclareMathOperator*{\minimize}{minimize}
\DeclareMathOperator*{\sto}{subject\; to}

\newcommand{\trace}{{\mbox{\textrm{Tr}}}}
\newcommand{\tr}{{\rm Tr}}

\newcommand{\bW}{{\mathbf{W}}}

\newcommand{\bv}{{\textbf{v}}}
\newcommand{\bV}{{\textbf{V}}}
\newcommand{\bH}{{\mathbf{H}}}

\newcommand{\bI}{\mathbf{I}}

\newcommand{\bU}{\mathbf{U}}
\newcommand{\bu}{\mathbf{u}}

\title{Cross Layer Provision of Future Cellular Networks}
\name{H.\ Baligh, M.\ Hong, W.-C.\ Liao, Z.-Q.\ Luo, M.\ Razaviyayn, M.\ Sanjabi, and R.\ Sun\address{}\thanks{This research is supported in part by
the NSF, Grant No.\ CCF-1216858, and in part by a research gift from
Huawei Technologies Inc.}\thanks{The authors' names are in alphabetical order.}}
\begin{document}
\maketitle
\begin{abstract}
{To cope with the growing demand for wireless data and to extend
service coverage, future 5G networks will increasingly
rely on the use of low powered nodes to support massive connectivity in diverse set of applications and services \cite{HuaweiSDN}. To
this end, virtualized and mass-scale cloud architectures are proposed as promising technologies for 5G in which all the nodes are connected via a backhaul network and managed centrally by such cloud centers. The significant computing power made available by the cloud technologies has enabled the implementation of sophisticated signal processing algorithms, especially by way of parallel processing, for both interference management and network provision. The latter two are among the major signal processing tasks for 5G due to increased level of frequency sharing, node density, interference and network congestion.}
This article outlines several theoretical and practical aspects of joint interference management and network provisioning for future 5G networks.
A cross-layer optimization framework is proposed for joint user admission, user-base station association, power control, user grouping, transceiver design as well as routing and flow control. {We show that many of these cross-layer tasks can be treated in a unified way and implemented in a parallel manner using an efficient algorithmic framework called WMMSE (Weighted MMSE).} Some recent developments in this area are highlighted and future research directions are identified.
\end{abstract}

\section{Introduction}
\begin{center}
 \begin{figure}[htb]
     {\includegraphics[width=
0.8\linewidth]{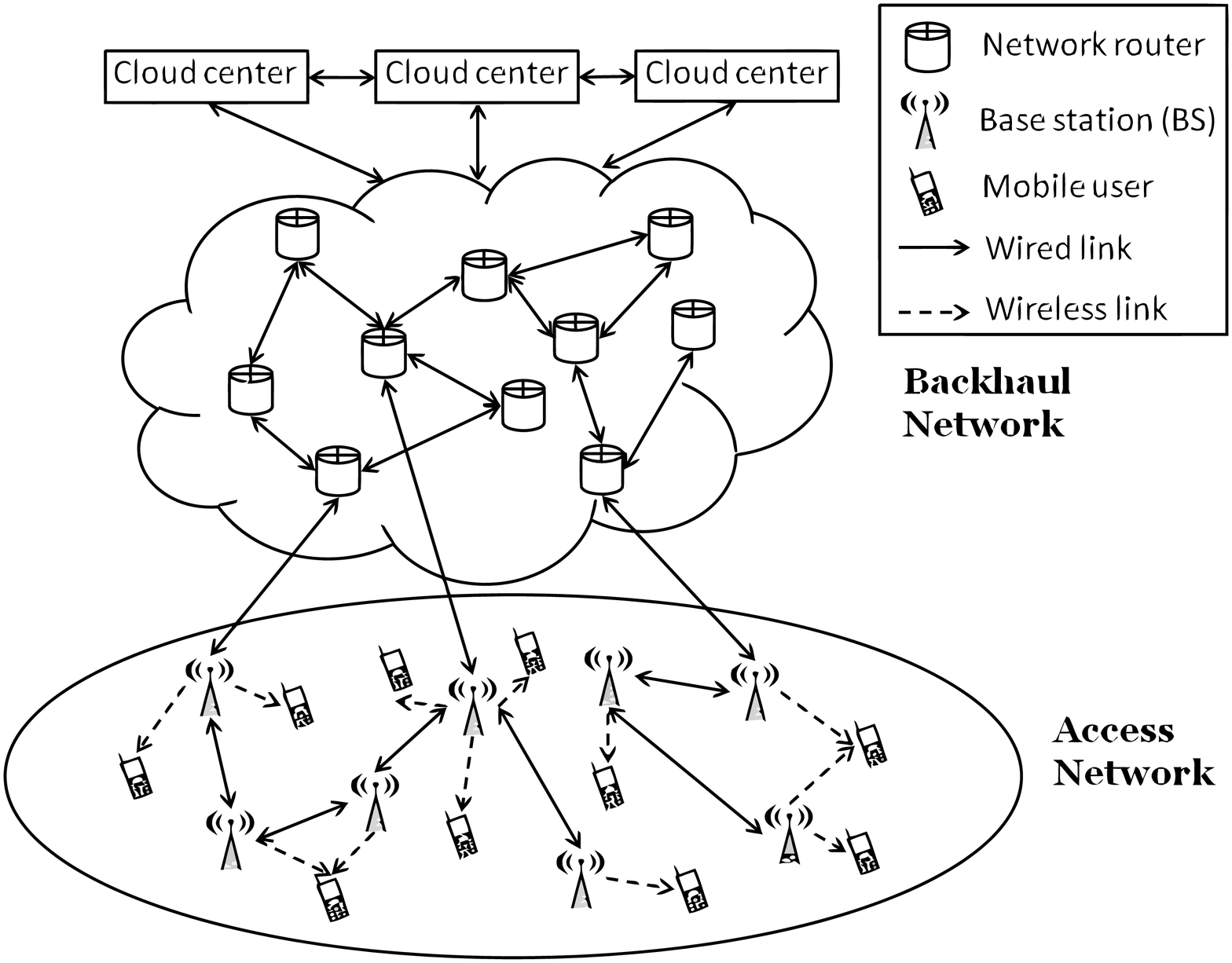}
\vspace*{-0.7cm}
\caption{\footnotesize Illustration of the SDN-RAN.}\label{figConfiguration} \vspace*{-0.4cm}}
\end{figure}
\end{center}

To increase network capacity and  spectral efficiency, network operators have been adding many low power micro/pico/femto base stations,  relays and WiFi access points (AP) to the network, thereby reducing the signal transmission distances.  This has resulted in the so-called heterogeneous network (HetNet) architecture; see \cite{damnjanovic2011survey} and the references therein. The HetNet architecture naturally replaces the traditional single hop access mode between the high-power base station and its users by a wireless mesh network consisting of a large number of densely deployed wireless access points with either wireline or wireless backhaul support; see Figure~\ref{figConfiguration}. {A natural extension to this concept would be virtualized radio access (VRA) and
mass-scale cloud architectures which proposed as promising technologies for 5G \cite{HuaweiSDN}. In VRA, all nodes are
connected via a backhaul network and managed centrally by
cloud centers.} Such a multi-hop network must be self-organized and fast adaptive to the changes of traffic demands caused by users joining, leaving, or requesting for new data packets at any time. To maximize the performance of such networks, the multiuser interference - which is a major performance limiting factor - should be astutely managed through advanced signal processing techniques.  In addition, traffic engineering within the entire radio access network (RAN) including the backhaul network should be jointly optimized with resource allocation across the wireless links.

 In this article, we present a cross-layer optimization framework for joint resource allocation and provision of future 5G networks. Our approach integrates several important cross-layer techniques: 1) advanced signal processing techniques for physical layer interference management; 2) MAC layer algorithms to handle user scheduling, base station assignment and base station clustering; 3) network layer solutions such as software defined networking (SDN) \cite{SDN,HuaweiSDN,ChinaMobileCRAN} to manage the network services and to control individual flows. Such cross-layer optimization is more challenging than the traditional cellular network optimization. Routing and scheduling of user traffic in the presence of multiuser interference is a difficult problem by itself. It is made more complicated by practical issues such as backhaul capacity limitations, channel state information (CSI) overhead, and distributed implementation of the algorithms. Our goal is to highlight some recent advances in this area, illustrate the potential of signal processing in the provision of future 5G networks and identify some future research directions.

{The concept of cross-layer optimization is not new and is known to be computationally complex. However, with the significant computing power at cloud centers of future 5G networks, we believe the time has come to consider the use of advanced signal processing algorithms for network provision. In fact, with the network densification and cloud RAN architecture, it is crucial to jointly provision the RAN and backhaul networks, addressing such issues as users-base station assignments, interference management, traffic engineering in a coordinated manner. Our proposed cross layer approach addresses many of these important issues in a coherent algorithmic framework based on a WMMSE technique \cite{christensen08,schmidt2009minimum,shi11WMMSE_TSP}.
}
\section{A Cross-layer Formulation}
\label{sec:format}

\subsection{{A General Model}}
For simplicity, we restrict our discussion to the downlink direction in which the traffic flows from the core network to the users through the wired/wireless backhaul network and the RAN.

Let the backhaul network be composed of a set $\mathcal{N}$ of routers and a set $\mathcal{L}^{\rm w}$ of wired links, whose main purpose is to route the traffic towards the RAN. Each backhaul link $l\in \mathcal{L}^{\rm w}$ has a fixed capacity, denoted as $\bar{C}^{l}$. The RAN consists of a set $\mathcal{Q}$ of Base Stations (BSs) and a set $\mathcal{I}$ of users. 
Assume that each BS/AP is equipped with $M$ transmit antennas, and that the total bandwidth is divided into $F$ frequency tones. 

Each user $i$ has $N$ receive antennas, and achieves an instantaneous rate of $R_{i}[t]$ for a given time instance $t$. Such rate is a function of a collection of system parameters, defined across various network layers. These parameters include the transmit strategies of the BSs (e.g., precoders, power control mechanism), the cooperation strategies among the BSs (e.g., the CoMP scheme), the routing decisions within the backhaul and so on. The network provision in this context amounts to achieving certain overall performance by optimizing, possibly in different time-scales, all the system parameters. Let $U(\cdot)$ denote a utility function that measures the system level performance.  We are interested in solving the following cross-layer system utility maximization problem
\begin{align}
\begin{split}
{\rm maximize}&\quad U\left(\{R_{i}[t]\}_{i\in \mathcal{I}}\right)\label{eq:generic}\\
\mbox{subject to}&\quad \mbox{Per BS resource constraints}\\
&\quad \mbox{Per data flow QoS constraints}\\
&\quad \mbox{Network structure constraints}\\
&\quad \mbox{Per node flow conservation constraints}.
\end{split}
\end{align}
At this stage the problem is described in a very generic form. Below we illustrate how each component of the problem can be instantiated in practice.

A large family of utility functions is the ``$\alpha$-fair"
utility functions, which is given by
\begin{align}
U\left(\{R_{i}[t]\}_{i\in\mathcal{I}}\right)=\sum_{i\in\mathcal{I}}\frac{(R_{i}[t])^{1-\alpha}}{1-\alpha}\label{eq:AlphaUtility}.
\end{align}
Different choices of the parameter $\alpha$ give different priorities to user fairness and overall system
performance. For example, when $\alpha=0$, we obtain the sum rate utility: $U(\{R_{i}[t]\}_{i\in\mathcal{I}})=\sum_{i\in\mathcal{I}}R_{i}[t]$. When $\alpha=1$ we obtain the proportional fair utility $U(\{R_{i}[t]\}_{i\in\mathcal{I}})=\sum_{i\in\mathcal{I}}\log(R_{i}[t])$. Popular choices of $\alpha$ and their corresponding utility functions can be found in \cite{BSassignmentMaz} and the references therein. Further, when certain parameters in the system (e.g., the channel realizations) are random, we can choose the average throughput $\sum_{i\in\mathcal{I}}\mathbb{E}\left[R_{i}[t]\right]$ as the system utility. In general, the system utilities expressed as a function of the users' rates are typically concave, but are nonconcave in terms of the system parameters (e.g., power, routing variables) due to multiuser interference \cite{hong12survey}.

The first type of constraints in \eqref{eq:generic} is related to the BSs' resources, or their individual transmit strategies. Suppose a BS $q$ uses a transmit precoder $\mathbf{V}^{q}_{i}[f,t]\in\mathbb{C}^{M\times N}$ to transmit to user $i$ on tone $f$ at time $t$. Then we have the following per-BS transmit power constraint
\begin{align}\label{eq:PowerConstraint}
\sum_{f=1}^{F}\sum_{i\in \mathcal{I}}\tr\left[{\mathbf{V}^{q}_{i}[f,t](\mathbf{V}^{q}_{i}[f,t])^H}\right]\le \bar{P}^{q}
\end{align}
where $\bar{P}^{q}$ is BS $q$'s power budget.  {\color{black}Other constraints in this category include per group of antenna constraints, user scheduling constraints, or the zero forcing constraints.}

The second type of constraints in \eqref{eq:generic} is related to the QoS required by the users. One common such requirement is given by $R_{i}[t]\ge \gamma_{i}[t],\; \forall\; i\in\mathcal{I}$,
where $\gamma_{i}[t]$ is the predefined minimum rate requested by user $i$ at time $t$. Alternatively, one can require that the outage probability is bounded above, i.e., $\mathbb{P}(R_{i}[t]\le \gamma_{i})\le \delta_{i},\; \forall\; i\in\mathcal{I}$.

The third type of constraints is closely related to the densification of the BS site deployment in 5G systems. New architectures such as the HetNet offer unprecedented flexibility in terms of how the access network can be formed. For example, we can cluster a subset of BSs to cooperatively serve a user $i$, and optimize the cluster membership to yield the best performance. To further elaborate on such network structure constraint, let us define a set of binary variables $\{z^{q}_{i}\}$ to represent the BS-user association ($z^{q}_{i}=1$ means BS $q$ serves user $i$, and zero otherwise). The set of {\it serving BSs} for user $i$ is then given by $\mathcal{S}_{i}:=\{q\mid q\in \mathcal{Q}, z^{q}_{i}=1\}$, and the set of {\it active user} for BS $q$ is $\mathcal{I}^q:=\{i\mid i\in\mathcal{I}, z^q_i=1\}$. Then constraining on $|\mathcal{S}_{i}|$ controls the size of each cooperative BS cluster. Other structural constraints include the ones that dictate that, at a given time, only a subset of BSs in $\mathcal{Q}$ are activated. Such constraints are useful to keep the operational costs of the access network under control.

The last type of constraints has to do with the control of data flow in the backhaul network. Without loss of generality, suppose that each user $i$ requests a single data flow. Then we use $s(i)$ and $d(i)$ to denote a source-destination pair for a given data flow, where $s(i)$ typically represents a Packet Data Gateway, and $d(i)$ is some user device. Let us use $\mathcal{L}^{\rm wl}$ to denote the set of wireless link in the system. That is, $\mathcal{L}^{\rm wl}\triangleq \{(s,d,f)\mid s\in\mathcal{Q},~
d\in\mathcal{I},~f=1\cdots, F\}$ with $(s,d,f)$ being the wireless link
from BS $s$ to user $d$ on tone $f$. Denote the rate for user $i$ on link $l$ as $R^{(l)}_{i}$. Then the following {\it flow conservation constraint} must hold true
\begin{align}
\begin{split}\label{eq:ConservationConst}
&\sum_{l\in{\rm In}(v)}R^{(l)}_{i}+1_{\{s(i)\}}(v)R_{i}=\sum_{l\in{\rm
Out}(v)}R^{(l)}_{i}\\
&\quad\quad+1_{\{d(i)\}}(v)R_{i}, \; \forall
\; i_k\in  \mathcal{I},~\forall\;
v\in\mathcal{V}
\end{split}
\end{align}
where ${\rm In}(v)$ and ${\rm Out}(v)$ denote the set of links going
into and coming out of a node $v$ respectively;  $1_{\mathcal{A}}(x)$ denotes the indicator
function for a set $\mathcal{A}$, i.e., $1_{\mathcal{A}}(x)=1$ if $x\in\mathcal{A}$, and
$1_{\mathcal{A}}(x)=0$ otherwise.

{\color{black}Finally, we give a concrete example showing how the users' rates are dependent on various system parameters introduced so far. Let $\mathbf{H}^{q}_i[f,t]\in\mathbb{C}^{N\times M}$ denote the channel matrix between BS $q$ and user $i$ on tone $f$ at slot $t$. Let $\sigma_i$ denote the noise power at user $i$. Further define a binary variable $\alpha_i[f,t]\in\{0,1\}$ to signify whether user $i$ is served on channel $f$ at slot $t$. Then user $i$'s rate $R_i[t]$, {under perfect CSI (Channel State Information),} can be expressed as\footnote{We use natural logarithm throughout this article.}
\begin{align}\label{eq:rate function}
{R}_i[t] & =\sum_{f=1}^{F}\alpha_{i}[t,f]\log \det \bigg(\bI + \sum_{q\in\mathcal{S}_i} \bH_i^{q}[t,f]\bV_{i}^{q}[t,f] \nonumber\\
&\quad\quad\times\sum_{q\in\mathcal{S}_i}(\bV_{i}^{q}[t,f])^H (\bH_i^{q}[t,f])^H   \mathbf{C}_i^{-1}[t,f]\bigg)
\end{align}
where $\mathbf{C}_i[t,f]\in\mathbb{C}^{N\times N}$ is the interference matrix for user $i$
\begin{align}
\mathbf{C}_i[t,f]=\sigma_{i}^2 \bI +  \sum_{(q,j): j \ne i} \bH_{i}^{q}[t,f]\bV_{j}^{q}[t,f] (\bV_{j}^{q}[t,f])^H (\bH_{i}^{q}[t,f])^H.\nonumber
\end{align}}

\subsection{{The Challenges and the Overview of Solutions}}
The generic problem \eqref{eq:generic} is of huge size, because of the increasingly large number of of access nodes in 5G systems as well as the fact that it covers quite a few aspects of the system design. It would be very challenging, if not impossible, to optimize \eqref{eq:generic} directly in its most general form. In practice, system parameters are usually optimized over different time scales. For example, network structure typically changes at a slower rate compared with the transmit/receive beamformer. Therefore at any given time instance, a reduced version of \eqref{eq:generic} is solved, maybe with much smaller problem dimension.

{ This article presents a few important cross-layer design problems formulated as special cases of problem \eqref{eq:generic}. We discuss existing approaches for these problems, and advocate a unifying framework based on the WMMSE (weighted MMSE) method. We demonstrate that the WMMSE \cite{christensen08,schmidt2009minimum,shi11WMMSE_TSP}, known as one of the state-of-the-art methods for physical layer precoder design, can be generalized in multiple ways to deal with cross-layer designs in 5G networks. More importantly, the resulting schemes are often amendable to efficient parallel implementation, which suits ideally to the cloud-based software-defined architecture of the 5G networks. }

%

\section{Cross-Layer Network Management}
\label{sec:pagestyle}
{
From a global perspective, the densification and heterogeneity of the access network give rise to new design issues across different layers of the network. In fact, the traditional boundary between different layers of the network has been blurred. For example, the possibility of using multiple closely located BSs to jointly serve one user impacts the traffic routing strategy in the network layer, which further affects the ways that the scheduling in the MAC layer and interference management in the physical layer are performed. In this section, we address various issues in the joint design across the physical, MAC and network layer. We take a bottom-up approach, and show how physical layer signal processing techniques can be generalized to deal with MAC layer problems such as scheduling and BS management/assignment, or network layer problems such as traffic management and congestion control. In particular, the well-known WMMSE algorithm \cite{christensen08,schmidt2009minimum,shi11WMMSE_TSP} is presented as a unifying framework for such purpose. 

}


\subsection{Physical layer interference management}
\label{sec:BF}
{Let us first consider the simple classical problem of precoder design in a multi-user wireless network, where only the physical layer transmit and receive strategies are the design variables. }
Assume the users have been properly scheduled in different time/frequency slots
and we only consider the problem in one time/frequency slot. Moreover, we assume that the user-BS association has been determined and each user is served by only one BS, i.e. the network is an IBC (Interfering Broadcast Channel).
The utility maximization problem in an IBC can be formulated as
\begin{equation}\label{eq:phy}
\begin{split}
{\rm maximize}&\quad U\left(\{R_{i}\}_{i\in \mathcal{I}}\right)\\
\sto&\quad \eqref{eq:rate function},\ \sum_{i\in \mathcal{I}^q} \tr\left[{\mathbf{V}^{q}_{i}(\mathbf{V}^{q}_{i})^H}\right]\le \bar{P}^{q}, \quad \forall q. \\
\end{split}
\end{equation}

{Except for some special cases (e.g. maximizing the minimum rate in the MISO or SIMO IC), problem (\ref{eq:phy}) is nonconvex and NP-hard (see \cite{hong12survey} and the references therein).  A few nonconvex instances of problem (\ref{eq:phy}), such as the sum rate maximization in MISO or SIMO IC, can be solved by global optimization
methods 
(see \cite{zheng2014maximizing} and the references therein).
 However, these algorithms are usually too complex to be used for large scale networks.  Moreover, it is not clear whether and how they can be extended to more general network settings such as MIMO IBC, or networks with flexible BS association and limited backhaul availability.
In general, finding the global maximum for problem (\ref{eq:phy}) is NP-hard and is difficult even for reasonably sized networks. 
Consequently, much research effort has been devoted to designing efficient algorithms that produce high quality suboptimal solutions to problem (\ref{eq:phy}); see \cite{bjornson13,hong12survey} for recent surveys. In this subsection, we briefly review the WMMSE algorithm as it forms the basis of the cross-layer optimization approach presented in this article.}

The WMMSE algorithm was first proposed in \cite{christensen08} for the MIMO broadcast channel and later extended to MIMO IC where each user transmits a single data stream \cite{schmidt2009minimum}. It was extended significantly to MIMO IC and MIMO IBC/I-MAC where each user transmits multiple data streams, and to a large family of utility functions including the sum rate utility \cite{shi11WMMSE_TSP}. The main idea is to transform the original sum utility maximization problem to a weighted MMSE minimization problem where the weights are adaptively updated. We illustrate this idea for the sum rate maximization in MIMO IC where each user only transmits a single data stream.
Suppose the transmit beamformer of BS $i$ is $\bm v^i \in \mathbb{C}^{M \times 1} $ (i.e. $\bV_i^i$ in the formulation \eqref{eq:phy}) and the receive beamformer of BS $i$ is $\bm u_i \in \mathbb{C}^{N \times 1}$ which has unit norm. The SINR of user $i$ is given by $\gamma_i = \frac{ | \bm u_i^H \bH_{i}^i \bm v^{i} |^2  }{ \sum_{j \neq i} | \bm u_i^H \bH_i^j \bm v^{j} |^2 + \sigma_i^2   }$, and the rate of user $i$ is given by $R_i = \log(1+\gamma_i)$. With these notations, the sum rate maximization problem becomes
\begin{equation}\label{eq:phy, sum rate of MIMO IC}
\begin{split}
\maximize_{\{ \bm u_i, \bm v^i\} } &\quad \sum_{i} R_i, \\
\sto&\quad \|\bm v^i \|^2 \leq \bar{P}^{i},\quad \forall \;i. \\
\end{split}
\end{equation}

The MSE (mean squared error) of user $i$ is given as
$ e_i  = |\bm u_i^H \bH_{i}^{i} \bm v^{i} - 1|^2 + \sum_{j\neq i} |\bm u_i^H \bH_{i}^{j} \bm v^j|^2 + \sigma_i^2 .$
There is a well-known relationship between SINR and MMSE:  MMSE = 1/(1+SINR). More precisely, for any given $\{\bm v^i \}$, we have
\begin{align}\label{eq:MSE_Rate}
1+ \max_{\bm u_i} \gamma_i = \max_{\bm u_i}  \frac{1}{e_i }.
\end{align}
Consequently, the sum rate maximization problem \eqref{eq:phy, sum rate of MIMO IC} is equivalent to the following problem  \cite{shi11WMMSE_TSP}:
\begin{equation}\label{eq:sum log e of MIMO IC}
\begin{split}
\minimize_{\{\bm u_i, \bm v^i\}} &\quad \sum_{i}\log( e_i ), \\
\sto&\quad \|\bm v^i \|^2 \leq \bar{P}^{i},\quad \forall\ i. \\
\end{split}
\end{equation}

 One can further prove that problem \eqref{eq:sum log e of MIMO IC} is equivalent to the following problem (in the sense that there is a one-to-one correspondence between the stationary points of the two problems)
\begin{equation}\label{eq:phy, after WMMSE transform}
\begin{split}
\minimize_{\{ \bm u_i, \bm v^i, w_i\} } &\quad  \sum_i ( w_i e_i - \log(w_i) ) \\
\sto&\quad \|\bm v^i \|^2 \leq \bar{P}^{i},\quad \forall\ i. \\
\end{split}
\end{equation}


Problem \eqref{eq:phy, after WMMSE transform} is an adaptively weighted MMSE problem and can be solved by alternate optimization.
In specific, $e_i$ is a convex quadratic function over $\{\bm u_i\}$ and $\{\bm v^i\}$ respectively, thus $\{\bm u_i\}$ and $\{\bm v^i\}$ can be updated in
closed forms (with an additional bisection step for updating $\{ \bm v^i \}$); the optimal $w_i$ is given by $w_i = 1/e_i$.
The convergence of WMMSE algorithm can be derived from the classical theory of the alternate optimization \cite{shi11WMMSE_TSP}.
{The details of WMMSE algorithm for sum rate maximization are given in Algorithm \ref{algo: WMMSE}.}
{We emphasize that the $\bm v$ subproblem can be solved very efficiently because of the equivalent transformation from \eqref{eq:phy, sum rate of MIMO IC} to \eqref{eq:phy, after WMMSE transform}, which is the key technique of WMMSE algorithm. As will be seen in subsequent sections, the alternate optimization framework and the simplification of the $\bm v$ subproblem are crucial to generalize WMMSE algorithm to solve cross-layer design problems.
}


\begin{algorithm}
\caption{WMMSE algorithm for sum rate maximization in MIMO IC (single beam case)}
\label{algo: WMMSE}
\begin{algorithmic}
\small
\STATE Initialize $\bm v^{i}, \bm u_{i}, w_i $  randomly.
\REPEAT
\STATE  $\bullet \;$ $\bm u_i  \leftarrow \left( \sum_{j=1}^K \bH_{i}^j  \bm v^{j}  (\bm v^j)^H (\bH_{i}^j)^H + \sigma_i^2 \bm I \right)^{-1} \bH_{i}^i \bm v^{i} , \forall i;  $
$\vspace{0.2cm}$
\STATE $\bullet\;$ $ w_i  \leftarrow (1 - \bm u_i^H \bH_{i}^i  \bm v^i )^{-1} , \; \forall i; $
$\vspace{0.2cm}$
\STATE $\bullet \;$ $\bm v^i  \leftarrow \left( \sum_{j=1}^K w_j (\bH_{j}^i)^H \bm u_{j} \bm u_j^H \bH_{j}^i + \mu_i^* \bm I \right)^{-1} \bH_{i}^i \bm u_{i} w_i , \; \forall i  $, \\
 where $\mu_i^*$ is computed by bisection such that $ \| \bm v^i  \|^2 \leq \bar{P}^i $.
\UNTIL Convergence
\end{algorithmic}
\end{algorithm}

 {Assuming all computation is executed at cloud centers, WMMSE algorithm can be easily implemented in a parallel fashion. Let $\bm u, \bm v, \bm w$ denote the sets of variables $\{\bm u_i\}, \{v^i \}, \{w_i \}$ respectively.
 Fixing the two sets of variables $\bm u$ and $\bm w$, the objective function of \eqref{eq:phy, after WMMSE transform} is decomposable across all BSs with respect to $\bm v$. As a result, the update of each $\bm v^i$ only depends on $\bm u$ and $\bm w$, and does not depend on other $\bm v^j, \forall j\neq i$. All $\bm v^i$ can be updated in a parallel fashion. Similarly, the variables $\bm u_i$ and $w_i$. can also be updated in parallel.}  

 The WMMSE algorithm can be interpreted as an inexact alternate optimization approach to solve problem \eqref{eq:sum log e of MIMO IC}, where $\{\bm u_k \}$ and $\{\bm v^k \}$ are updated iteratively. The receive beamformers $\{\bm u_k \}$ are updated to MMSE receivers, which exactly minimize $\sum_k \log(e_k)$.
  The transmit precoders $\{\bm v^k \}$ are updated to minimize  $\sum_j \frac{\partial \log(e_j)}{\partial e_j} e_j$, a ``linear'' approximation of $\sum_j \log(e_j)$, where $\frac{\partial \log(e_j)}{\partial e_j} = 1/e_j$ (i.e. the optimal $w_j$ for \eqref{eq:phy, after WMMSE transform}) is computed based on the previous iterates. With this interpretation, the convergence of WMMSE algorithm can also be derived from the result in \cite{Razaviyayn12SUM}.

{
We summarize the advantages of the WMMSE algorithm as follows.
 First, the alternate optimization framework and the simplicity of each subproblem makes WMMSE algorithm easily extendable to cross-layer design. Additionally, no stepsize tuning is needed. Second, the WMMSE algorithm exploits the special structure of the rate function, thus converges faster than general nonlinear optimization methods such as the gradient descent algorithm.
Third, since at each step the subproblem can be decomposed across each user, it is amenable to parallel and distributed implementation.

Let us briefly compare the WMMSE algorithm with the parallel SCA (Successive Convex Approximation) algorithm \cite{scutari13decomposition}, which is a variant of the parallel version of the interference pricing algorithm \cite{schmidt2009distributed}.
Similar to the WMMSE algorithm, the parallel SCA exploits the structure of the rate function, can be implemented in parallel, and has been extended to solve some cross-layer design problem \cite{sardellitti2013joint}. One of the major differences is that the parallel SCA aims to design transmit covariance matrices rather than the beamformer, so it cannot be directly used when the number of transmit antennas are larger than the receive antennas, or there are restrictions on the number of transmitted data streams. Since parallel SCA shows good performance in certain problem setup, it will be interesting to see whether it can be generalized to other cross-layer design problems.

}   

{
\subsection{Joint PHY-MAC layer optimizations: scheduling and resource allocation}}
\label{sec:BFschdlngClstrng}
{Once we know how to deal with physical layer precoder design using WMMSE, we move up one layer and show that it is in fact easy to  integrate various tasks from the MAC layer, such as user scheduling across different time slots/frequency tones, BS assignment and BS clustering. The resulting PHY-MAC joint designs are again stepsize free, decompose nicely across the network nodes, and are thus easily parallelizable. }

\subsubsection{Beamforming and scheduling}
For a HetNet, user scheduling and beamforming are two effective techniques to mitigate multi-user interference. Joint user scheduling and beamforming has been investigated in different works including \cite{GroupingTSP, yu2011multicell}. The basic  problem is to design linear transmit/receive beamformers and schedule the users across a fixed set of time slots/frequency tones in a way that maximizes a given system utility. For simplicity of presentation, let there be one frequency tone and $T$ time slots. We assume that the BS assignment is fixed and each user is served by only one BS, i.e., $|\mathcal{S}_i| = 1, \;\forall i \in \mathcal{I}$. {For notational simplicity, we only present the algorithm for the simple interference channel model, although the algorithm and the analysis can be generalized to the interfering broadcast channel \cite{GroupingTSP}.} Let us define a binary variable $\alpha_i[t] \in \{0,1\}$ to indicate if user~$i$ is served in time slot~$t$ or not. We assume no joint decoding across different time slots and therefore the rate of each user is the summation of the rates over different time slots/frequency tones. {Notice that the time slots here are much longer than the symbol length and typically of the size of the block used for channel coding.} Making this {assumption}, the joint beamforming and scheduling problem can be formulated as
\small
{
\begin{align}
\maximize_{{\mathbf{V},\alpha}}&\quad U\left(\{{R}_i\}_{i\in \mathcal{I}}\right)\label{eq:scheduling}\\
\sto&\quad\tr \left[{\mathbf{V}^{i}[t](\mathbf{V}^{i}[t])^H} \right]\le \bar{P}^{i}, \;\forall\ i, t \nonumber\\
&\quad {R}_i = \sum_t \alpha_{i}[t]\log \det \bigg(\bI +  \bH_i^{i}\bV^{i}[t] (\bV^{i}[t])^H (\bH_i^{i})^H  \times \nonumber\\
&\quad\quad\bigg( \sigma_{i}^2 \bI +  \sum_{j \neq i} \bH_{i}^{j}\bV^{j}[t] (\bV^{j}[t])^H (\bH_{i}^{j})^H
\bigg)^{-1} \bigg). \nonumber\\
&\quad\alpha_i[t] \in \{0,1\},\forall\ i, t \nonumber
\end{align}
}
\normalsize
{It can be shown that without loosing optimality, the scheduling variables $\alpha_i[t]$ can be initially set to 1. Furthermore, local linearization of the utility function around the current rate values results in similar problem as in the sum rate maximization problem. Using this observation, an algorithm similar to WMMSE (Algorithm~\ref{fig:pseudo_code_MIMO_group}) can be used to solve the optimization problem \eqref{eq:scheduling}; see \cite{GroupingTSP} for more details. After solving the above optimization problem, we can update the scheduling variables $\alpha_i[t]$ using the obtained precoders.}

\begin{algorithm}[h]
\caption{Joint scheduling and beamforming algorithm}
\label{fig:pseudo_code_MIMO_group}
\begin{algorithmic}
\small
{
\STATE initialize $\bV^{i} [t]$'s randomly
\REPEAT
\STATE  $\bullet \;$ $\bU_{i}[t] \leftarrow \left({\sum_{j} \bH_{i}^{j}\bV^{j}[t]
    (\bV^{j})^H (\bH_{i}^{j})^H+\sigma_{i}^2\bI}\right)^{-1}\times$ $\bH_{i}^{i}\bV^{i}[t]$, $\;\; \forall \;i ,t$
$\vspace{0.2cm}$
\STATE $\bullet\;$ Update $\bW_i[t]$ by calculating the local gradient of the utility function; see \cite{GroupingTSP} for details
$\vspace{0.2cm}$
\STATE $\bullet\;$ $\bV^{i}[t] \leftarrow \left({\sum_{j} (\bH_{j}^{i})^H
    \bU_{j}[t]\bW_{j}[t](\bU_{j}[t])^H\bH_{j}^{i} }+ \mu_{i}^* \bI\right)^{-1}$ $\times(\bH_{i}^{i})^H
    \bU_{i}[t] \bW_{i}[t]$, $\forall \;i,t$
\UNTIL convergence
\STATE \textbf{if} $\|\bV^{i}[t]\|>\epsilon$
\textbf{then} $\alpha_i[t]\leftarrow 1$;
\textbf{otherwise} $\alpha_i[t]\leftarrow 0,\;\forall i,t$
}
\end{algorithmic}
\end{algorithm}

\begin{figure}[htbp]
\centering
\includegraphics[width=3.5in]{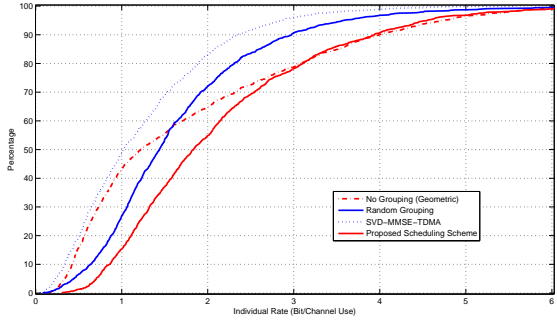}
\caption{Rate CDF of different methods \cite{GroupingTSP}}
\label{fig:RK57M4N2G3NM5Er01}
\end{figure}

\normalsize
Figure~\ref{fig:RK57M4N2G3NM5Er01} illustrates the performance gain of the joint scheduling and beamforming optimization in a 19-hexagonal wrap-around cell layout. There are 285 users served by 57 BSs in the network. The other simulation details can be found in \cite{GroupingTSP}. The figure shows the achieved rate cumulative distribution function (CDF) of different approaches: The ``No Scheduling" corresponds to the WMMSE algorithm with no scheduling; The ``Random Scheduling" curve represents a random scheduler combined with WMMSE algorithm; While in the ``SVD-MMSE-TDMA" approach, each base station serves its own users in a TDMA fashion by using the SVD (singular value decomposition) precoder and the users deploy MMSE receivers. As shown in Figure~\ref{fig:RK57M4N2G3NM5Er01}, the joint beamforming and scheduling can significantly improve the system throughput as well as the user fairness.

\subsubsection{Beamforming and BS assignment }
{ The densification of the BS deployment gives the users the flexibility of choosing  its serving BS from potentially a large pool of nearby BSs. Traditionally, BS assignment is made on the basis of signal strength (or the distances between the BSs and a given user).  However such greedy scheme often leads to suboptimal solutions and  causes unfairness among users, especially when the network is congested \cite{stridh2006system}. It may be more beneficial to assign users to a different BS when the closest BS is congested \cite{stridh2006system}. Such freedom in assignment, if properly exploited, can result in substantial gains both in terms of network throughput and user fairness \cite{BSassignmentMaz,bjornson2013massive}.}

{We show below how to integrate BS assignment with beamforming in the WMMSE framework}. Consider the following problem over both the BS assignment and beamforming variables $\{z_i^q,\bV_i^q\}_{i,q}$:
\begin{align}
\maximize_{z,\bV} &\quad {U}\left(\{R_i\}_{i\in \mathcal{I}}\right) \nonumber\\[-5pt]
\sto & \quad \sum_q z_i^q \leq 1, \quad z_i^q \in \{0,1\},\;\forall i,q \label{eq:Assignmentconstraint}\\[-5pt]
& \quad\sum_{i \in \mathcal{I}^q}  {\rm Tr} \left(\bV_i^q (\bV_i^q)^H\right) \leq \bar{P}^q \nonumber\\[-5pt]
&\quad R_i =  \sum_q z_i^q \log \det \bigg(\bI + \bH_i^q \bV_i^q (\bV_i^q)^H (\bH_i^q)^H \bigg(\sigma_i^2\bI\nonumber\\[-5pt]
&\quad\quad+ \sum_{(j,p)\neq(i,q)} \bH_i^p \bV_j^p (\bV_j^p)^H (\bH_i^p)^H\bigg)^{-1}\bigg) \nonumber.
\end{align}

\begin{figure}[htbp]
\centering
\includegraphics[width=3.5in]{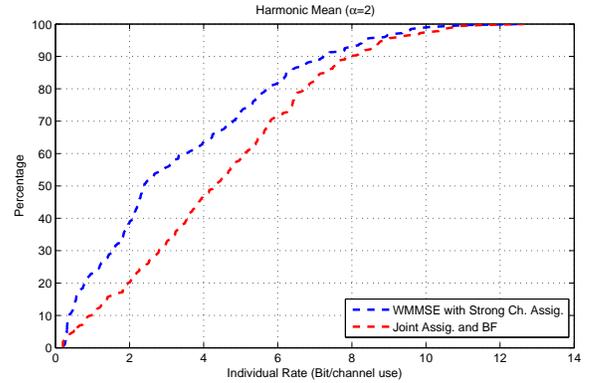}
\caption{Rate CDF achieved by joint BS assignment and beamforming \cite{BSassignmentMaz}}
\label{fig:M4N4L16P110exp2new2}
\end{figure}

If the optimization variable $\{z_i^q\}$ is fixed, then the above optimization problem reduces to  \eqref{eq:phy}. The extra binary constraint in \eqref{eq:Assignmentconstraint} makes the problem difficult to solve.
One way to alleviate this difficulty is to relax it to $0\leq z_i^q\leq 1$. {It is not hard to see that the relaxation is tight when the objective is linearized with respect to the rate of the users. On the other hand, when the association variables are fixed and the objective is linearized with respect to the rate of the users, the problem is similar to the beamforming for sum rate maximization (cf.\ problem \eqref{eq:phy, sum rate of MIMO IC}), which can be handled using the WMMSE algorithm. In order to update the association variables, we use gradient projection method. Combining the gradient projection step for updating the association variables and the WMMSE algorithm for updating the beamformers, we can solve the above problem to a stationary point;  see \cite{BSassignmentMaz} for more details.} Figure~\ref{fig:M4N4L16P110exp2new2} illustrates the performance gain achieved by the joint precoder design and BS assignment. For benchmark, we have chosen the greedy BS assignment method, i.e., first each user is assigned to the BS with the strongest channel and then the WMMSE is used to optimize the beamformers. Figure~\ref{fig:M4N4L16P110exp2new2} shows that the joint beamforming and BS assignment achieves significantly higher throughput and fairness. {For other relaxations of the BS assignment problem, we refer the readers to \cite{stridh2006system}.}

\subsubsection{BS clustering for CoMP}
\label{sec:clusteringCoMP}
{When there are a large number of BSs available, coordinated transmission and reception is shown to be very effective in improving the overall spectrum efficiency \cite{foschini2006coordinating}. Two popular approaches are coordinated beamforming and joint processing.} In the first approach, a user's data resides only at its serving BS and the beamformers are jointly optimized among the coordinated BSs to suppress  multi-user interference. The second approach is joint transmission where  the user data signals are shared among the cooperating BSs, resulting in a single virtual BS with a large set of antennas. In this way, the HetNet is reduced to a large MIMO broadcast channel. However, full cooperation among all BSs may require significant overhead across the backhaul. In practice, it is desirable to limit the amount of coordination by grouping the BSs into cooperating clusters of small sizes, within which joint processing is performed. In this way, user's data signals are shared only within clusters, thus reducing the overall backhaul signaling overhead. Many recent works have developed various BS clustering strategies for such purpose where clusters are formed either greedily or by an exhaustive search procedure. Once the clusters are formed, various approaches can be used to design beamforming strategies for each BS.

\begin{figure}[htbp]
\centering
\vspace{-0.2cm}
\includegraphics[width=2.5in]{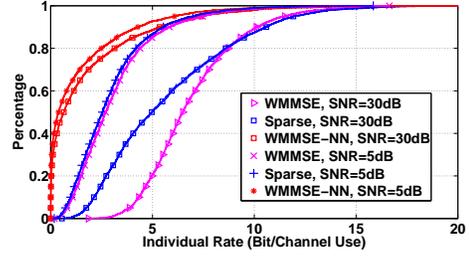}
\caption{Rate CDF achieved by joint BS clustering and beamforming \cite{hong2013joint}}
\label{fig:SWMMSE}
\vspace{-0.3cm}
\end{figure}

{An important cross-layer design issue in this context is how to form BS clusters in conjunction with beamforming and BS coordination so as to strike the best tradeoff among system throughput performance and signalling overhead. To formally state the problem, let us define $\mathcal{Q}_i$ to be a subset of BSs that can potentially cooperate to serve a particular user~$i$, e.g., the set of BSs deployed in the same cell as user~$i$. The goal of clustering is to reduce the system overhead by setting the less beneficial precoders to zero. Such sparse precoder structure can be imposed by solving the following problem, where a group Lasso regularizer is introduced in the objective to promote sparsity among potential precoders \cite{hong2013joint}}:
\begin{align}\label{eq:sparse}
\begin{split}
\maximize_{\{\bV_i^q\}} &\quad {U}\left(\{R_i\}_{i\in \mathcal{I}}\right)  - \lambda \sum_{i \in \mathcal{I}} \sum_{q \in \mathcal{Q}_i} {\|\bV_i^q\|_F}\\
 \sto& \quad \eqref{eq:rate function},\ \sum_{i \in \mathcal{I}^q}  {\rm Tr} \left(\bV_i^q (\bV_i^q)^H\right) \leq \bar{P}^q,\;\forall \ q
\end{split}
\end{align}
{ Once again the WMMSE algorithm is naturally suited to solve this problem. The key here is to recognize that after performing the equivalent transformation (cf.\ \eqref{eq:phy, after WMMSE transform}), the $\mathbf{u}$, $\mathbf{w}$ subproblems are exactly the same as before (with closed-form solutions), and the $\mathbf{v}$ subproblem becomes a quadratic problem penalized by a group-LASSO regularizer}, which is still easy to solve. By again using the alternate optimization approach, problem \eqref{eq:sparse} can be solved to a stationary point; see \cite{hong2013joint} for more details. {It is worth noticing the tradeoff in the choice of parameter $\lambda$: larger values of $\lambda$ result in smaller size clusters, but lower system throughput. Based on the numerical experiments, a simple rule for  selecting the value of $\lambda$ is suggested in \cite{hong2013joint}.} Figure~\ref{fig:SWMMSE} shows the performance gain of the sparse-WMMSE algorithm.  This numerical experiment is run over a network with 20 BSs and 40 users in the system; the number of transmit (resp.\ receive) antenna is 4 (resp.\ 2). The algorithm is evaluated in the low and high SNR regime; see \cite{hong2013joint} for more details. For comparison, we have also simulated the performance of two other approaches: WMMSE and NN-WMMSE. In the WMMSE, full cooperation is considered among the BSs of each cell and the WMMSE algorithm is used to optimize the precoders. Clearly, this corresponds to one single cluster which requires most system overhead. In contrast, the NN-WMMSE only picks the BS with the strongest channel to serve each user. Thus, the NN-WMMSE method corresponds to greedy BS assignment followed by optimal beamforming.

{
\subsection{Joint physical-network layer optimization: resource allocation and traffic engineering}\label{sec:SDN}}
In light of growing traffic demand and large network size, backhaul links may be  capacity limited \cite{HuaweiSDN,SDN,ChinaMobileCRAN}.  Therefore traffic engineering within the multi-hop backhaul network (e.g., traffic routing) must be considered together with resource allocation in the radio air interface (e.g., precoder design and scheduling).

One interesting approach to such joint optimization problem, which is gaining support from both academia and industry, is to manage the entire network by a few cloud centers. The shift of the computation tasks from a large number of heterogeneous BSs to a few cloud centers is attractive to the operators, as it allows for an effective and energy efficient way of managing the entire network. This new architecture is called cloud-RAN or SDN-RAN \cite{HuaweiSDN,SDN,ChinaMobileCRAN}.

There are two main methods in the literature to deal with the per-link capacity constraints in the backhaul. The first method allows the cloud center to compute a joint precoding strategy for all the BSs, and then compress the precoded messages before sending to the BSs via the backhaul; see \cite{Park13} and the references therein. The limited backhaul capacity determines the level of compression needed for each data stream. To illustrate the idea, let us consider a joint downlink compression and sum-rate maximization problem in the cloud-RAN. Suppose that each BS $q$ is connected to the cloud center via a dedicated line with capacity $\bar{C}^q$. Let $\mathbf{V}^q\triangleq\{\mathbf{V}^q_{i}\}_{i\in\mathcal{I}}$ denote the precoder used by BS $q$ to serve all the users, and let $\mathbf{H}_i\triangleq\{\mathbf{H}^{q}_i\}_{q\in\mathcal{Q}}$ denote the channel for user $i$. Define $\mathbf{V}_i$ similarly. Let $\mathbf{\Omega}_{q,p}\in\mathbb{C}^{M\times M}$ denote the correlation matrix between the quantization noises of BS $q$ and $p$. Let $\mathbf{\Omega}$ be the compression covariance with $\mathbf{\Omega}_{q,p}$ as its $(q,p)$th block component. By using the precoding and compression scheme described in \cite{Park13}, user $i$'s rate can be expressed as
\begin{align}
R_i(\mathbf{V},\mathbf{\Omega})&=\log\big|\mathbf{I}+\mathbf{H}_i(\mathbf{V}\mathbf{V}^H+\mathbf{\Omega})\mathbf{H}^H_i\big|\nonumber\\
&-\bigg|\mathbf{I}+\mathbf{H}_i(\sum_{j\ne i}\mathbf{V}_j\mathbf{V}_j^H+\mathbf{\Omega})\mathbf{H}^H_i\bigg|.
\end{align}
Then the problem can be formulated as
\begin{subequations}
\begin{align}
\maximize_{\mathbf{V},\mathbf{\Omega}\succeq 0}& \quad \sum_{i\in\mathcal{I}}R_i(\mathbf{V}, \mathbf{\Omega})\label{CloudRAN}\\
\sto&\quad \sum_{q\in\mathcal{S}}\log\frac{|\mathbf{V}^q\mathbf{V}^q+\mathbf{\Omega}_{q,q}|}{|\mathbf{\Omega}_{q,q}|}\le \sum_{q\in\cal{S}}{\bar{C}^q}, \;\forall\; \cal{S}\subseteq\cal{Q},\label{eq:Compression}\\
&\quad \trace\left[ \mathbf{V}^q(\mathbf{V}^q)^H+\mathbf{\Omega}_{q,q}\right]\le \bar{P}^{q}, \;\forall\; q\in \mathcal{Q},
\end{align}
\end{subequations}
where the constraints in \eqref{eq:Compression} ensure that the messages can be reliably transferred to the BSs through the backhaul. With this formulation, the compression covariance as well as transmit precoders can be jointly designed using optimization algorithms such as successive convex approximation \cite{scutari13decomposition,Razaviyayn12SUM}. However, this line of work usually assume that there is a single-hop direct connection between the BSs and the cloud centers, and that the routing of the traffic within the backhaul has been predetermined.

The second approach combines the traditional utility based traffic engineering with the resource allocation in access network. The idea is to optimize the routing scheme in the backhaul and the power allocation/precoding strategy in the access network together. Different from the precoding-compression scheme, this approach allows for flexible backhaul structure, and the resulting algorithm can be implemented in parallel among the cloud centers. To illustrate the main ideas, let us consider an instance of such joint optimization problem for a single time slot (thus ignoring the time indices):
\begin{subequations}\label{MaxMinSDN}
\begin{align}
\maximize&\quad {U}\left(\{R_i\}_{i\in \mathcal{I}}\right)\\
\sto&\quad R_{i}^{(l)}\geq
0,~i\in\mathcal{I},~\forall\;l\in\mathcal{L},\mbox{~and~}
\eqref{eq:PowerConstraint} \label{eq:RateNonnegative}\\
&\quad\sum_{i\in\mathcal{I}}R_{i}^{(l)}\leq \bar C^{l},\;\forall\ l\in\mathcal{L}^{\rm w}, ~\mbox{and}~\eqref{eq:ConservationConst}  \label{eq:CapacityConstraintsWired}\\
&\quad\sum_{i\in\mathcal{I}}R_{i}^{(l)}\leq \bar R^{l},\;\forall\ l\in\mathcal{L}^{\rm wl}\label{eq:CapacityConstraintsWireless},
\end{align}
\end{subequations}
where $\bar{R}^{l}$ represents the capacity of a wireless link $l$, which is a function of transmit strategies of all BSs interfering link $l$. Compared with the problems studied in the previous sections, the joint design in \eqref{MaxMinSDN} is more difficult because of the large number of additional variables $\{R^{(l)}_i\}$ and constraints \eqref{eq:RateNonnegative}-\eqref{eq:CapacityConstraintsWired}, which together describe how the traffic is routed within the backhaul.

Variants of the joint optimization problem \eqref{MaxMinSDN} have been studied extensively in the framework of cross-layer optimization, see \cite{Chiang07}. However, most of the existing methods assume that the air interface is {\it interference free}, leading to a tractable convex problem \cite{Chiang07}.  As argued throughout this article, such interference-free transmission is unrealistic in future RANs. 


Recently, \cite{Liao13SDN} proposes to solve the joint routing and interference management problem by combining the WMMSE and the powerful Alternating Direction Method of Multipliers (ADMM). The ADMM algorithm is designed for structured convex optimization problem with linearly coupled constraints.
The idea in \cite{Liao13SDN} is quite simple: use WMMSE to take care of the interference management in the RAN, and apply ADMM to handle the traffic engineering in the backhaul.
%
%
%
%
%
%
%
To highlight ideas, we adopt the max-min utility function and consider the single antenna case (generalization to MIMO case and other utilities is nontrivial, but possible). {Transmit and receive beamformers now become scalars, denoted using lower case letters.} Similar to the WMMSE algorithm, we first transform problem \eqref{MaxMinSDN} into an equivalent but easily manageable form. To this end, introduce two sets of additional optimization variables ${\bf
u}\triangleq\{u_{l}\mid l\in\mathcal{L}^{\rm wl}\}$ (the receive beamformers) and ${\bf
w}\triangleq\{w_{l}\mid l\in\mathcal{L}^{\rm wl}\}$ (the weights), and explore the well-known
rate-MSE relationship for each wireless link $l=(s,d,f)\in\mathcal{L}^{\rm wl}$:{
\begin{align}\label{Rate-MSERelation}
\bar R^{l}=\max_{u_{l}, w_{l}}
c_{1,l}+c_{2,l}v^{s}_{d}[f]-\hspace{-0.35cm}\sum_{n=(s',d',f)}\hspace{-0.35cm}c_{3,ln}|v^{s'}_{d'}[f]|^{2}.
\end{align}}
Here $(c_{1,l}, c_{2,l}, c_{3,ln})$ are functions of $\bf u$ and $\bf w$ given by $
c_{1,l}=1+\log(w_{l})-w_{l}(1+\sigma_{d}^2|u_{l}|^{2})$,
{$c_{2,l}=2w_{l} {\rm Re}\{{u_{l}^{*}}h_{d}^{s}[f]\}$, and $
c_{3,ln}=w_{l}|u_{l}|^{2}|h_{d}^{s'}[f]|^{2}$.}
The difficult wireless channel rate constraints \eqref{eq:CapacityConstraintsWireless} are then replaced with the following {\it Rate-MSE} constraint{
\begin{align}
&\sum_{i\in\mathcal{I}}R_{i}^{(l)}\leq c_{1,l}+c_{2,l}v^{s}_{d}[f]-\hspace{-0.35cm}\sum_{n=(s',d',f)}\hspace{-0.35cm}c_{3,ln}|v^{s'}_{d'}[f]|^{2},\;\forall\;l\in\mathcal{L}^{\rm wl}\label{RateMSEConstraint}.
\end{align}}%
This transformation is motivated by the following facts: {\it 1)} any {$\{\bf v, \bf R\}$} satisfies \eqref{RateMSEConstraint} must also satisfy \eqref{eq:CapacityConstraintsWireless}; {\it 2)} constraint \eqref{eq:CapacityConstraintsWireless} is replaced by \eqref{RateMSEConstraint}, so the modified version of  problem \eqref{MaxMinSDN}, is convex with respect to any single variable $\mathbf{w}$, $\bf u$ and {$\{\bf v, \bf R\}$}. Therefore, the alternating optimization technique is again applied to solve the transformed problem; see the N-MaxMin Algorithm outlined { in Algorithm \ref{algo: N-MaxMin}}.
As shown in \cite{Liao13SDN}, the iterates generated by this algorithm converge to a stationary solution of problem \eqref{MaxMinSDN}.


\begin{algorithm}
\caption{{ N-MaxMin algorithm for minimum rate maximization in SDN-RAN (SISO case)}}
\label{algo: N-MaxMin}{
\begin{algorithmic}
\small
\STATE Initialize feasible $\bf u$, $\bf w$, $\bf v$, and $\bf R$ randomly
\REPEAT
\STATE  $\bullet$ $\forall\;l=(s,d,f)\in\mathcal{L}^{wl}$
\begin{align*}
u_{l}&\leftarrow(\sum_{n=(s',d',f)}\mid h_{d}^{s'}[f]|^{2}|v_{d'}^{s'}[f]|^{2}+\sigma_{d}^{2})^{-1}h_{d}^{s}[f]v_{d}^{s}[f]\\
w_{l}&\leftarrow (1-(h_{d}^{s}[f]v_{d}^{s}[f])^{*}u_{l})^{-1}
\end{align*}
\STATE $\bullet$ Update {$\{\bf v, \bf R\}$} via inner ADMM algorithm, cf. \cite{Liao13SDN}
\UNTIL Convergence
\end{algorithmic}}
\end{algorithm}

In the N-MaxMin, solving $\bf u$ and $\bf w$ is again easy and in closed-form. What is different from the sum-rate WMMSE algorithm so far is that the update for {$\{\bf v, \bf R\}$} is not in closed form. This step requires solving a huge convex optimization subproblem, which couples all the nodes in the system.
To make the entire algorithm scalable and suitable for distributed implementation, the ADMM is then used to decompose the huge problem for {$\{\bf v, \bf R\}$} into a collection of easy subproblems.
The trick here is to decouple the flow conservation constrains \eqref{eq:ConservationConst} and the rate-MSE constraints \eqref{RateMSEConstraint} by introducing a few extra variables. The interested readers are referred to \cite{Liao13SDN} for detailed description of the algorithm. We mention that all the steps of the resulting ADMM iterations are separable over the nodes/links of the network, and each of them can be updated in (semi)closed-form. More importantly, such separability structure makes the N-MaxMin algorithm implementable in parallel by a few cloud centers, each handling a subset of nodes/flows.

\begin{figure}[t]
\begin{center}
{ \subfigure[] []{\resizebox{.22\textwidth}{!}{\includegraphics{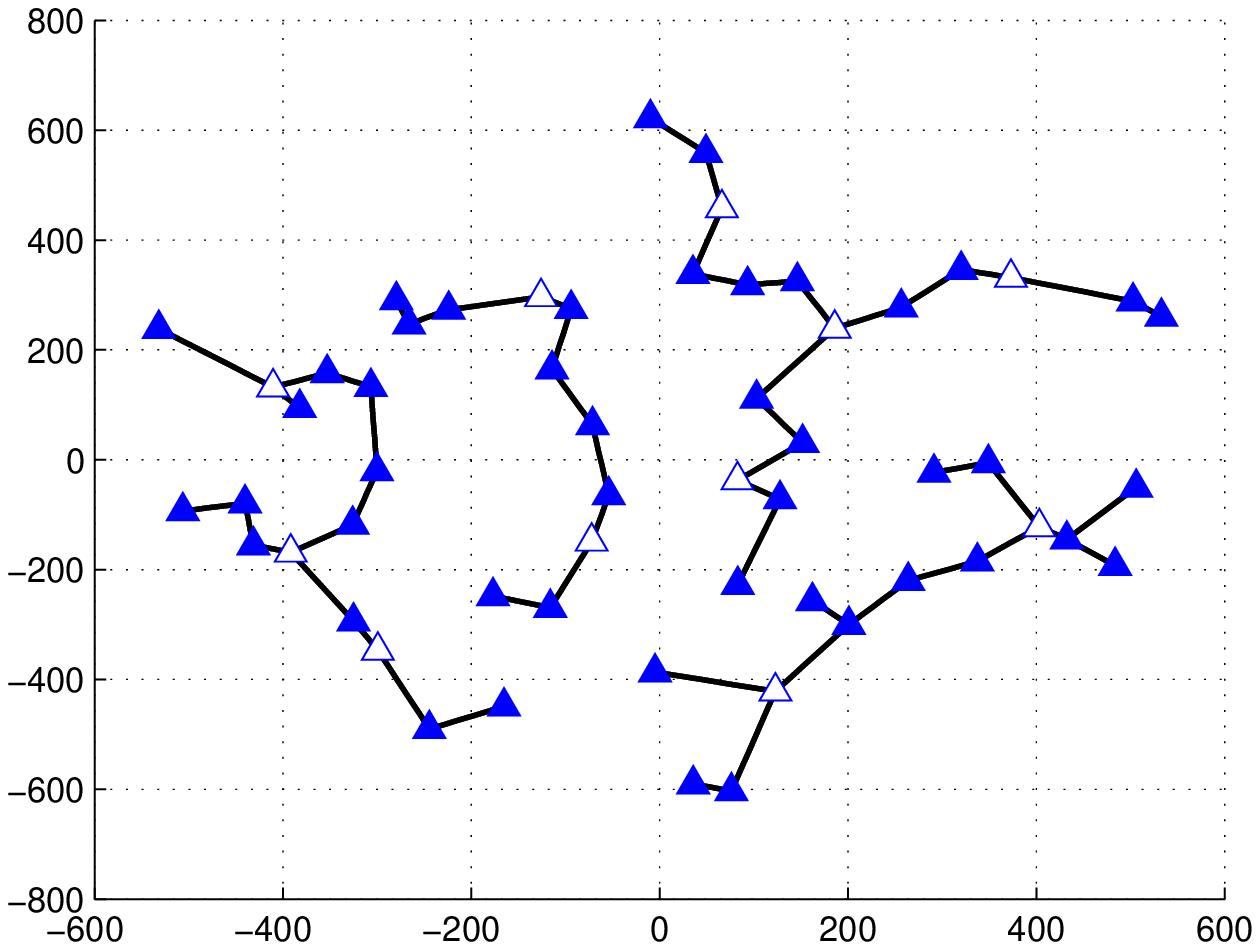}}}}
\hspace{1pc} { \subfigure[][ ]{\resizebox{.22\textwidth}{!}{\includegraphics{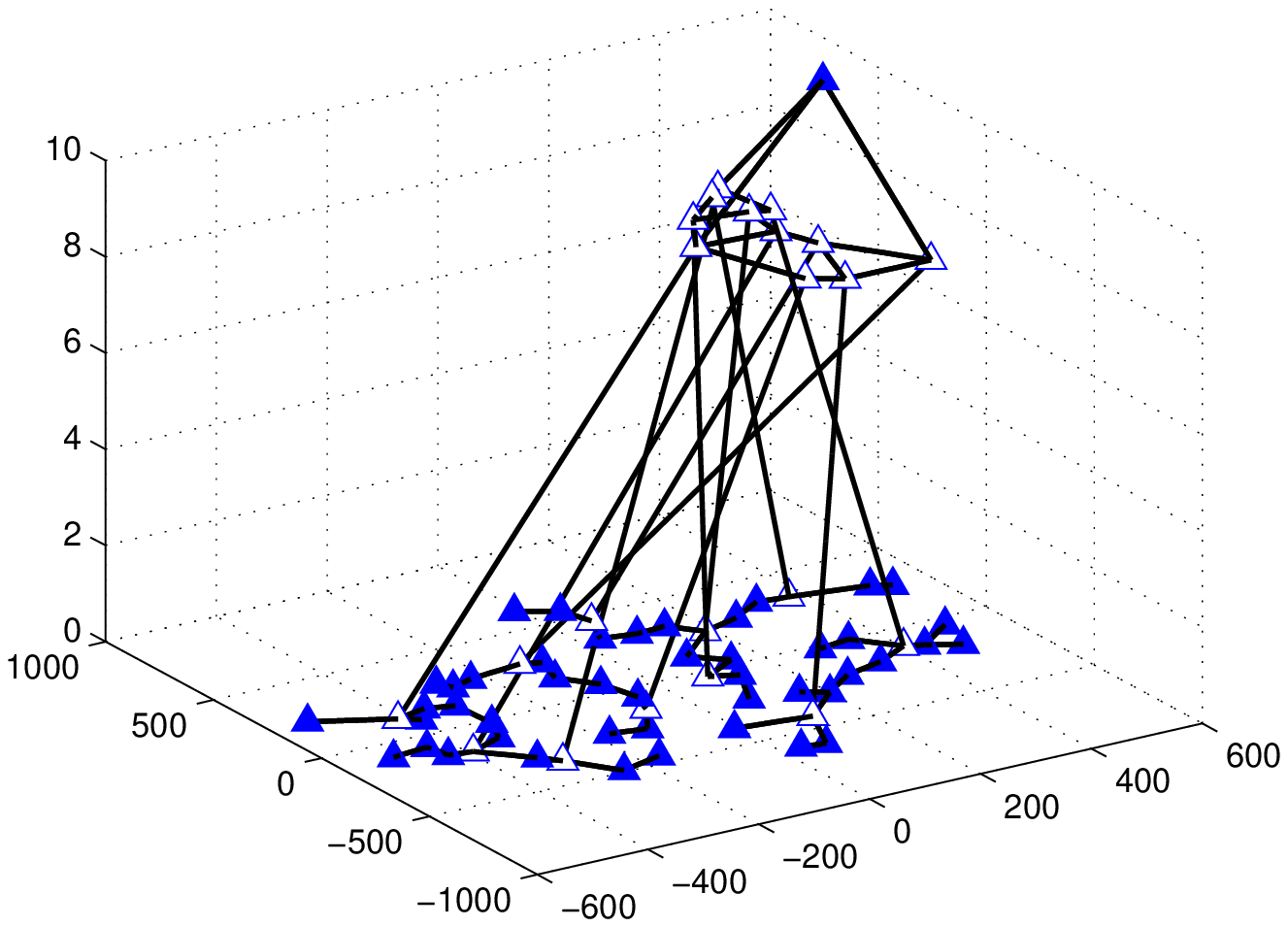}}
}}
\end{center}
\caption{\footnotesize The considered network topology. (a) The locations and the connectivity
of all the BSs.
(b) The connections between BSs and
routers, which are displayed in the upper part of the graph.
}\label{FigTopology}
\end{figure}

{\bf Numerical Examples:} Let us illustrate the performance of the N-MaxMin Algorithm using a few examples. The topology and the connectivity of the testing network are shown in Fig.\ \ref{FigTopology}. The source (destination) node of each commodity is randomly selected from network routers (mobile users). For the backhaul links of this network, a fixed capacity is assumed in the range between 2.88Mnats/s to 1.44Gnats/s, and they are the same in both directions. The number of subchannels is $K=3$ and each subchannel has $1$ MHz bandwidth, and the wireless links
follow the distribution $CN(0,(200/{\rm dist})^3)$, where
${\rm dist}$ is the distance between BS and mobile user.

\begin{figure}
\begin{center}
{ \subfigure[]
[]{\resizebox{.215\textwidth}{!}{\includegraphics{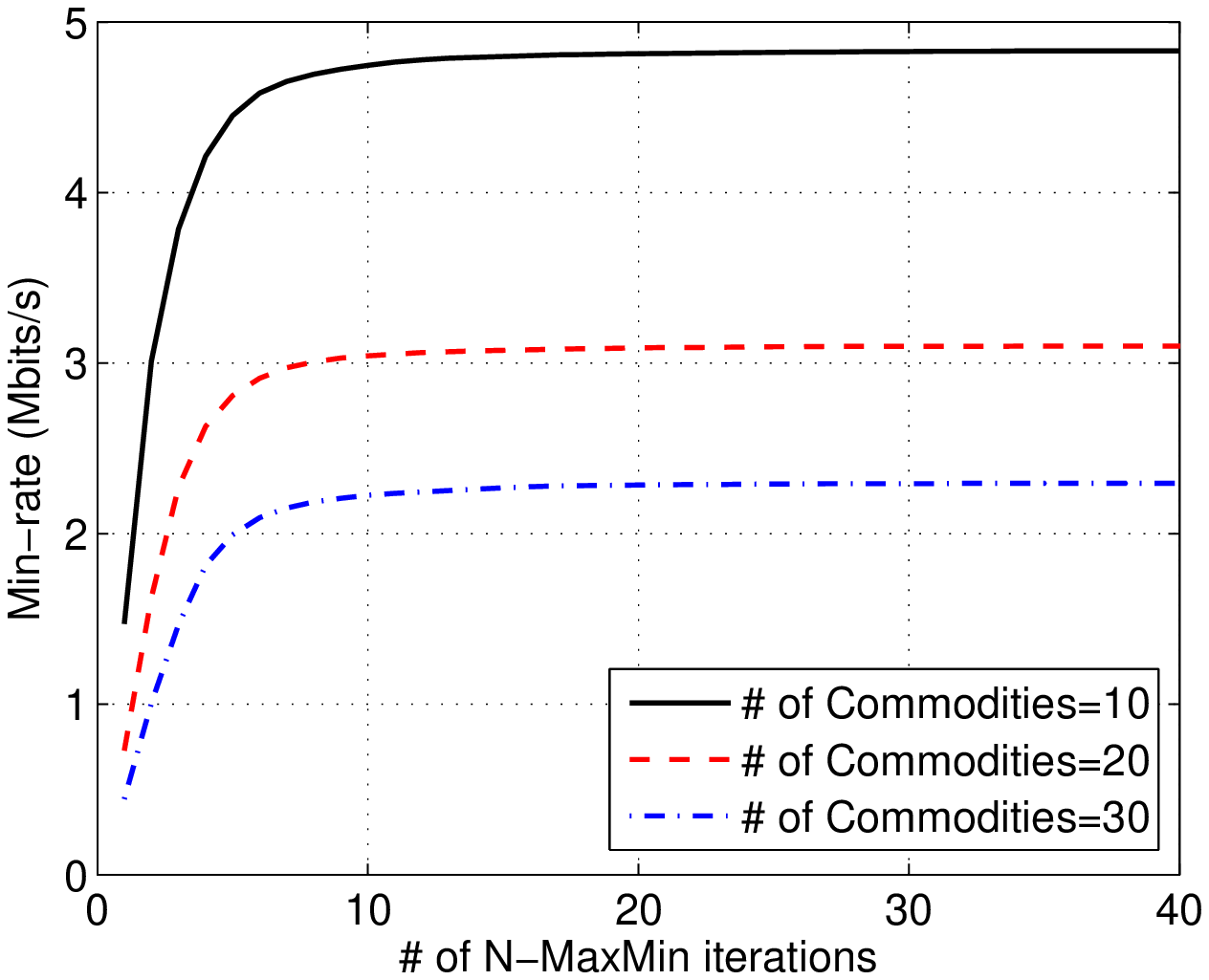}}}}
\hspace{1pc} { \subfigure[][
]{\resizebox{.225\textwidth}{!}{\includegraphics{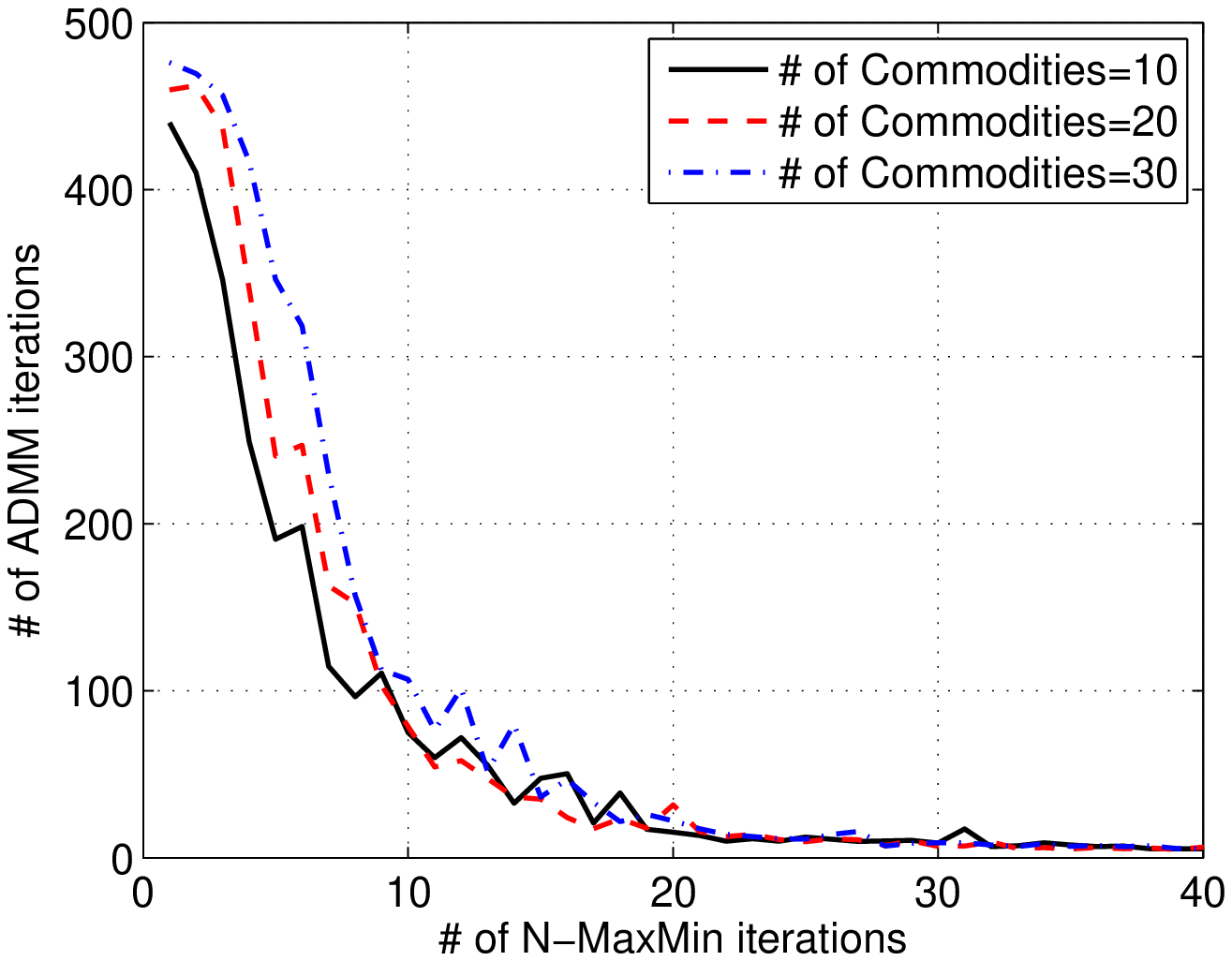}}
}}
\end{center}\vspace{-0.3cm}
\caption{\footnotesize The min-rate performance and the required
number of iterations for the proposed N-MaxMin algorithm. Part (a) plots
the obtained min-rate versus the iterations of N-MaxMin.
Part (b) plots the required number of inner ADMM iterations
v.s.\ the iteration for the outer N-MaxMin
algorithm. \cite{Liao13SDN}}\label{FigRateIte}\vspace{-0.3cm}
\end{figure}

To accelerate the initial steps, a maximum of $500$ inner iterations are allowed for the first $5$ outer iterations. Fig.\ \ref{FigRateIte} shows that the min-rate converges at about the $10$th outer iteration when
the number of commodities is up to $30$, while fewer than $500$ inner
iterations are needed per outer iteration. Moreover, after the
$10$th outer iteration, the number of inner ADMM iterations reaches
below $100$.

Next we demonstrate how parallel implementation can speed up the algorithm considerably. To illustrate the benefit of parallelization, we consider a larger network which is derived by merging two identical
networks shown in Fig.\ \ref{FigTopology} (a). For simplicity, we removed all the wireless links, so this reduces problem \eqref{MaxMinSDN} to a network flow problem (a very large linear program). Here, the N-MaxMin Algorithm is implemented by the Open MPI package with $9$ computation cores, and is compared  with Gurobi (a state of the art commercial LP solver) in terms of efficiency. In Table
\ref{TableTimeCompare}, we observe that parallel implementation leads to more
than 5 fold improvement in computation time, compared with its sequential counterpart. We also note that
when the problem size increases to  $8\times 10^4$ variables and over $10^5$ constraints, Gurobi becomes
slower than the proposed algorithm (implemented in parallel). Thus, the proposed ADMM algorithm scales well with problem size.

\begin{table}[t]
\begin{center}
\begin{tabular}{|c|c|c|c|c|c|}
\hline {\bf $\begin{array}{c}\mbox{\# of Commodities} \end{array}$} & {\bf 50} & {\bf 100} & {\bf 200} & {\bf 300} \\
\hline
{\bf $\begin{array}{c}\mbox{\# of} ~
\mbox{Variables} \end{array}$({\bf $\times 10^4$})} &
{\bf 1.4} & {\bf 2.9} & {\bf 5.8} & {\bf 8.7}\\
\hline
{\bf $\begin{array}{c}\mbox{\# of}~
\mbox{Constraints} \end{array}$({\bf $\times 10^4$})} & {\bf 2.1} & {\bf 4.2} & {\bf 8.4} & {\bf 13} \\
\hline
\hline
{\bf $\begin{array}{c}\mbox{Sequential Update (s) } \end{array}$} & 1.04 & 2.03 & 4.73 & 8.53 \\
\hline
{\bf $\begin{array}{c}\mbox{Parallel Update (s)} \end{array}$} & 0.20 & 0.37 & 0.75 & 1.10 \\
\hline
{\bf $\begin{array}{c}\mbox{Gurobi (s)} \\
 \end{array}$} & 0.20 & 0.64 & 1.65 & 2.51 \\
\hline
\end{tabular}
\end{center}
\caption{Comparison of computation time used by different implementations of the ADMM algorithm for the routing problem.
}\label{TableTimeCompare}
\vspace{-0.5cm}
\end{table}

\subsection{Practical considerations: overhead reduction }
\label{ssec:PracticalCons}
In the previous sections, the channel states are assumed known. In practice, the direct channel state is estimated using pilot messages. Performing only few channel estimation might lead to effects such as channel aging, resulting in imperfect channel estimation. For future cellular networks, allowing simultaneous use of frequency slots leads to interference. Hence, in addition to the direct channels, the states of the interfering links also need to be estimated. Moreover, all these channel estimations should be fed back to the transmitters. With the large number of co-channel transmissions, estimating the states of all interfering links is infeasible. A major challenge to implement the cross layer management algorithms (such as the ones presented in the previous sections) is the overhead associated with channel estimation. In this section we discuss some of the possible ways to partially relax the CSIT (channel state information at the transmitter side).

For wireless links, the channel state can be modeled as a random variable with a known distribution. The distributions can be calibrated using experimental data and are usually defined by long term physical parameters such as path-loss coefficients, propagation distance and channel training time/power. The first approach to CSIT mismatch is the use of robust optimization techniques; see e.g.,  \cite{DavidsonShenoudaRobust}. These methods are mostly designed for the worst case scenarios and therefore, due to their nature, are suboptimal when the worst cases happen with small probability. On the one hand, these methods can give guarantees on the quality of service constraint with high probability (measured based on the channel distribution). On the other hand, as they need to work with outage probability type of constraints, these methods are typically rather complex compared to their non-robust counterparts. In addition, they have a limited scope and cannot deal with realistic channel distributions due to analytical intractability. In fact, most of them use a Gaussian channel distribution with estimated channels as the means for the Gaussian distributions. Therefore, utilizing these methods still requires estimating the channel for all the links in the network (including all the interfering links), albeit the channel estimation need not be very accurate. For a large HetNet, this approach is still  not practical since estimating all the channel states requires excessive training overhead.

An alternative approach is to maximize the expected performance using a stochastic optimization framework which requires only the statistical channel knowledge rather than the full instantaneous CSI. This method is proposed in \cite{razaviyayn2013stochasticJournal} where {\color{black}the goal is to maximize the sum of expected utilities of the users. To illustrate, we can consider a similar formulation as \eqref{eq:phy, sum rate of MIMO IC} in an IC, but using an averaged sum rate as the objective (extensions to interfering broadcast channel (IBC) or CoMP is possible)
\begin{align}
\maximize_{\bv^{i}, \bu_i}~&~\sum_{i}\mathbb{E}[R_{i}]\label{eq:stochasticP}\\
\mbox{subject to}~&~ \|\bv^i\|\leq \bar{P}^i,\;\forall\; i.\nonumber
\end{align}
The algorithm can be viewed as a generalization of the WMMSE algorithm (Algorithm~\ref{algo: WMMSE}) to the stochastic setting. 
The algorithm proceeds by drawing{/receiving} a sample of the channels in each step using the known channel distributions, and then applying a step of the WMMSE to an  average of the utilities defined by the current as well as the  previous channel samples. The details of the algorithm are presented in Algorithm \ref{algo: SWMMSE}. As it can be seen that the steps of the algorithm are similar to Algorithm \ref{algo: WMMSE}, with the only difference being in the use of matrices $\mathbf{A}$ and $\mathbf{B}$ which serve to accumulate the information of all the {previous} channel samples thus far.}
{\color{black}
\begin{algorithm}
\caption{Stochastic WMMSE Algorithm}
\label{algo: SWMMSE}
\begin{algorithmic}
\STATE Initialize $\bv$ randomly such that $\|\bv^i\|^2=\bar{P}^i,\;\forall\; i$
\REPEAT
\STATE Obtain the new channel estimate/realization $\bH$
\STATE   $\bu_{i} \leftarrow \left(\sum_{j} \bH_{i}^{j}\bv^{j} (\bv^{j})^H (\bH_{i}^{j})^H+\sigma_{i}^2\bI\right)^{-1}\bH_{i}^{i}\bv^i,\; \forall\; i$
\STATE  $w_{i} \leftarrow \left(\bI-\bu_{i}^H\bH_{i}^{i}\bv^{i}\right)^{-1}, \;\forall\; i$
\STATE $\mathbf{A}_i \leftarrow \mathbf{A}_i + \sum_{j}w_{j}(\bH_{j}^{i})^H \bu_{j}\bu_{j}^H \bH_{j}^{i},\;\forall\; i $
\STATE $\mathbf{B}_i \leftarrow \mathbf{B}_i + (\bH_{i}^{i})^H \bu_{i} w_{i},\;\forall\; i$
\STATE $\bv^i \leftarrow (\mathbf{A}_i+ \mu_i^{*} \bI)^{-1} \mathbf{B}_i,\;\forall\; i$\\
 where $\mu_i^{*}$ is computed by bisection to ensure that $ \|\bv^i\|^2 \leq \bar{P}^i $.
\UNTIL convergence
\end{algorithmic}
\end{algorithm}
}
{Note that in stochastic WMMSE, transmit precoders $\bv$ are adapted to channels statistics known at transmitters, while receive beamformers are adapted optimally to real time channels. Therefore, the transmit precoder designed by Stochastic WMMSE algorithm is less sensitive to the changes in real time channel. Furthermore, in practice it is easy to incorporate the changes in statistical models of the channels into the algorithm by using a forgetting factor that reduces the effect of the past channel realizations on the choice of current precoder.}

We emphasize that the Stochastic WMMSE algorithm converges {to}
a stationary solution for any channel distributions (and not just Gaussian).
In practice, each user can estimate the direct channel and as
well as a few (say 1-3) interfering channels. The estimated channels
can be modeled as Gaussian random variables whose variances
may be specified by the training SNR. For the rest of interfering
channels, we can use standard path loss statistics plus scattering to
model them. Estimating path loss, which is slow varying, is much
easier and requires significantly less communication overhead than
estimating the fast changing channels. {Note that the Stochastic WMMSE is computationally as cheap as its non-stochastic counterpart of Section \ref{sec:BF}.}
{Moreover, it can be generalized to handle joint clustering and beamforming for CoMP transmission. Such generalization can be done easily by adding a sparsity promoting penalty term to the objective in \eqref{eq:stochasticP}. }

\begin{figure}
  \centering
    \includegraphics[width=0.45\textwidth]{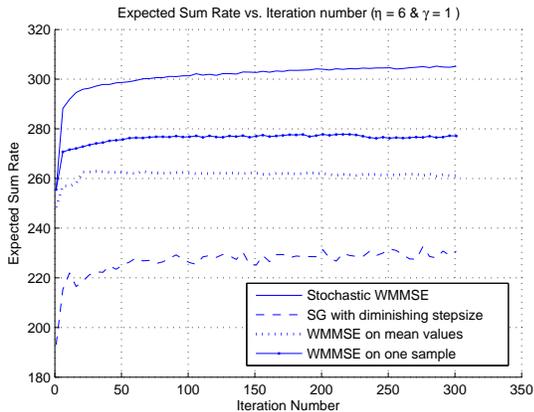}
      \caption{Expected sum rate vs. iteration number. We set $\eta = 6$, $\gamma = 1$ and consequently only 3\% of the channel matrices are estimated, while the rest are generated by their path loss coefficients plus Rayleigh fading. The signal to noise ratio is set ${\rm SNR}= 15$ (dB) \cite{razaviyayn2013stochasticJournal}.}\label{fig: Sim1}
      \vspace{-0.5cm}
\end{figure}

To evaluate the effectiveness of the Stochastic WMMSE algorithm,
we consider a network of 19 cells, each with 3 separate sectors,
for a total of $K = 57$ base stations. Each BS is equipped with
4 antennas and serves the users (all equipped with 2 antennas) in its
own sector. The path loss and the power budget of the transmitters
are generated using the 3GPP (TR 36.814) evaluation methodology
\cite{3gpp}. We assume that partial channel state information is available
for some of the links. In particular, each user estimates only
its direct channel in addition to the interfering links whose powers
are at most $\eta$ (dB) below its direct channel power ($\eta$ is a simulation
parameter). For these estimated channels, we assume a channel estimation
error model in the form of $\hat{h} = h + z$, where $h$ is the actual
channel; $\hat{h}$ is the estimated channel, and $z$ is the estimation error.
Given a MMSE channel estimate $\hat{h}$, we can determine the distribution
of $h$ as $\mathcal{CN} (\hat{h}, \frac{\sigma_{l}^{2}}{1+\gamma {\rm SNR}})$
 where $\gamma$ is the effective signal to noise
ratio (SNR) coefficient depending on the system parameters (e.g. the
number of pilot symbols used for channel estimation) and $\sigma_l$ is the
path loss. Moreover, for the channels which are not estimated, we
assume the availability of estimates of the path loss $\sigma_l$ and use them
to construct statistical models (Rayleigh fading is considered on top
of the path loss).

We compare the performance of four different algorithms:
{\it one-sample WMMSE, mean WMMSE, stochastic gradient}, and {\it Stochastic
WMMSE}. In the one-sample WMMSE and the mean WMMSE,
we apply the WMMSE algorithm \cite{shi11WMMSE_TSP} to one realization of all channels
and to the mean channel matrices, respectively. For the stochastic
gradient method, we use a diminishing step size rule to the ergodic
sum rate maximization problem. For more numerical experiments and the details of the  simulations, the interested readers are referred to \cite{razaviyayn2013stochasticJournal}. Figure \ref{fig: Sim1} shows
our simulation results when each user only estimates the strongest
3\% of its channels, while the others are generated synthetically according
to the channel distributions. The expected sum rate in each
iteration is approximated in this figure by a Monte-Carlo averaging
over 500 independent channel realizations. As can be seen from Figure
8, the Stochastic WMMSE algorithm significantly outperforms
the rest of the algorithms. 

\section{Concluding remarks}

A key feature of future wireless networks is the large number of base
stations, each with varying capability and connected to a backhaul
network to serve the user needs. How to best manage such networks
to meet the users' high speed, high mobility requirement is a major
challenge. In this article, we outline a cross layer approach, one that
leverages resource allocation in the physical layer, user scheduling in
the access layer and traffic engineering in the network layer to maximize
an overall system utility. Optimization and signal processing
are the key elements in this approach. 

{ While we have highlighted a few recent developments in this area, much remains to be done,
especially in the directions of high performance and low energy footprint network provision
algorithms that can be implemented efficiently in a parallel manner. In particular, it will be important to include the emerging concept of ``green communication"  \cite{Hasan11} into the algorithm design. Energy efficiency can be indirectly achieved by, for example, MAC layer mechanisms similar to the BS clustering in Section \ref{sec:clusteringCoMP}. That is, by dynamically {\it shutting down} a few BSs that is under-utilized, the overall energy consumption can be greatly reduced; see for example the recent work \cite{liao13admm}. Alternatively, physical layer approaches can be used to directly maximize the energy efficiency measured by the throughput per unit of energy consumption, see \cite{He14} and the references therein. It will be interesting to see how these separate approaches can be integrated in a cross-layer design framework. Another important direction is to evaluate and extend the algorithms proposed so far using realistic network layer protocols, traffic patterns and network models. Performance metrics such as throughput, delay, traffic queue stability, and achieved quality of service need to be carefully evaluated to validate the benefits brought by the cross-layer design.


}

\end{document}